\documentclass[preprint,11pt]{emulateapj}  
\usepackage{color}
\usepackage{epsfig}
\usepackage{subfigure,verbatim}

\newcommand{\unit}[1]{\nobreak{\mathrm{\;#1}}} 

\newcommand{\ditto}[1]{{_{\scriptsize{\rm{#1}}}}} 

\newcommand{\be}{\begin{eqnarray}}
\newcommand{\ee}{\end{eqnarray}}
\newcommand{\eq}[1]{Equation~(\ref{eq:#1})}
\newcommand{\eqn}[1]{(\ref{eq:#1})}
\newcommand{\fig}[1]{Figure~\ref{fig:#1}}

\def\simless{\mathbin{\lower 3pt\hbox
{$\rlap{\raise 5pt\hbox{$\char'074$}}\mathchar"7218$}}}   
\def\simmore{\mathbin{\lower 3pt\hbox
{$\rlap{\raise 5pt\hbox{$\char'076$}}\mathchar"7218$}}}   

\def\epsb{\epsilon_{\scriptsize{\rm B},-2}}

\begin{document}

\title{A Late-Time Flattening of Light Curves in Gamma-Ray Burst Afterglows}

\author{Lorenzo Sironi$^{1,3}$ and Dimitrios Giannios$^{2}$}
\affil{
$^{1}$Harvard-Smithsonian Center for Astrophysics, 60 Garden Street, Cambridge, MA 02138, USA\\
$^{2}$Department of Physics, Purdue University, 525 Northwestern
Avenue, West Lafayette, IN 47907, USA\\
$^{3}$NASA Einstein Postdoctoral Fellow}

\email{Email: lsironi@cfa.harvard.edu; dgiannio@purdue.edu.}

\begin{abstract}
The afterglow emission from Gamma--Ray Bursts (GRBs) is usually interpreted as synchrotron radiation from relativistic electrons accelerated at the GRB external shock, that decelerates from ultra-relativistic to non-relativistic speeds as it sweeps up the  surrounding medium. We investigate the temporal decay of the  emission from GRB afterglows at late times, when the bulk of the shock-accelerated electrons are non-relativistic. For a uniform circumburst medium, we show that such ``deep Newtonian phase'' begins at $t\ditto{DN}\sim 3\,\epsilon_{e,-1}^{5/6}t\ditto{ST}$, where $t\ditto{ST}$ marks the transition of the blast wave to the non-relativistic spherically-symmetric Sedov--Taylor solution, and $\epsilon_e=0.1\,\epsilon_{e,-1}$ quantifies the amount of shock energy transferred to the post-shock electrons. For typical parameters, the  deep Newtonian stage starts $\sim 0.5-$several years after the GRB. The radio flux in this phase decays as $F_{\nu}\propto t^{-3(p+1)/10}\propto t^{-(0.9\div1.2)}$, for a power-law distribution of shock-accelerated electrons with slope $2<p<3$. This is shallower than the commonly assumed scaling $F_\nu\propto t^{-3(5p-7)/10}\propto t^{-(0.9\div2.4)}$ derived by \citet{frail00}, which only applies if the GRB shock is non-relativistic, but the electron distribution still peaks at ultra-relativistic energies (a regime that is relevant for a narrow time interval, and only if $t\ditto{DN}\gtrsim t\ditto{ST}$, or $\epsilon_{e}\gtrsim0.03$). We discuss how the deep Newtonian phase can be reliably used for GRB calorimetry, and we comment on the good detection prospects of trans-relativistic blast waves at $0.1\div10\,$GHz with EVLA and LOFAR.
\end{abstract}

\section{Introduction}
Afterglow radiation from Gamma--Ray Bursts (GRBs) is attributed to external shocks produced by the interaction between the ultra-relativistic ejecta and the circumburst medium. Synchrotron emission from the shock-accelerated electrons powers the observed afterglow, detected in the $\gamma$-ray, X--ray, optical and radio bands \citep[e.g.,][]{sari98,wijers99,panaitescu00,granot02,kumar09}.
Nowadays, GRB afterglows are observed at times as early as the prompt emission at X-ray and optical frequencies, and up to 
a few years after the GRB at radio wavelengths.

As  the blast wave sweeps up the surrounding medium, the external shock slows down to non-relativistic velocities, and it finally approaches spherical symmetry \citep[e.g.,][]{sari98, sari99b, rhoads99, livio_waxman00, zhang09}. At this point, the dynamics can be described by the non-relativistic spherically-symmetric Sedov--Taylor (ST) solution. The radiation from the external shock can be calculated by assuming synchrotron emitting electrons with a power-law energy spectrum $dN/d\gamma\propto \gamma^{-p}$, for $\gamma\geq\gamma_{m}$.\footnote{For the sake of simplicity, in the following we assume that all of the post-shock electrons are contained in the power-law tail (i.e., the  number fraction of shock-accelerated electrons is $\zeta_e=1$). Also, throughout the text we reserve the lower case $\gamma$ for the random particle Lorentz factor and upper case $\Gamma$ for the fluid bulk motion.} The minimum Lorentz factor $\gamma_{m}$ is related to the fraction $\epsilon_e$ of shock energy transferred to the accelerated electrons by $\gamma_{m}-1=\frac{p-2}{p-1}\,\epsilon_e\, (\Gamma-1) \,m_p/m_e$, where $\Gamma$ is the bulk Lorentz factor of the blast wave. By assuming $\gamma_m\gg1$, \citet{frail00} found that the synchrotron flux in the radio band should decay as $F_\nu\propto t^{-3(5p-7)/10}$, after the shock becomes non-relativistic (i.e., $\Gamma-1\ll1$).

The assumption that the low-energy end of the electron distribution stays ultra-relativistic breaks down as soon as $\Gamma-1\lesssim \epsilon_e^{-1}m_e/m_p$. When the peak of the electron distribution drops down to non-relativistic energies, the system will transition to a new regime, which we shall call the ``deep Newtonian phase.'' In the deep Newtonian stage, the  theory of diffusive shock acceleration predicts that, while most of the shock-heated electrons are non-relativistic, the bulk of the electron energy is contributed by mildly relativistic particles with $\gamma_{\rm pk}\sim2$ \citep[e.g.,][]{bell_78, blandford_ostriker_78, blandford_eichler_87}. This is opposite to the  case typically considered in relativistic shocks, where for $\gamma_m\gg1$ and $p>2$ the particles with $\gamma_{\rm pk}\sim\gamma_m$ dominate {\it both} the number and the energy census \citep{sari98}.

In this work, we study the synchrotron light curves expected in the deep Newtonian regime. As we show in Section \ref{results}, we find that this phase begins at $t\ditto{DN}\sim 3\,\epsilon_{e,-1}^{5/6}t\ditto{ST}$, where $t\ditto{ST}$ marks the transition of the blast wave to the non-relativistic spherically-symmetric ST solution, and $\epsilon_{e}=0.1\,\epsilon_{e,-1}$. For $2<p<3$, we find that the radio flux decays as $F_{\nu}\propto t^{-3(p+1)/10}\propto t^{-(0.9\div1.2)}$, which is shallower than the regime discussed by \citet{frail00}, and nearly independent of the uncertain power-law slope $p$. The GRB afterglows with the longest follow-ups (e.g., GRB 030329, still visible in radio waves years after the explosion)  are likely to be in the deep Newtonian phase, so our results can be used to perform reliable calorimetric estimates of the explosions. Although our main focus is on GRB afterglows, our results can be equally applied to radio afterglows from tidal disruption events, radio supernovae, and trans-relativistic ejecta from neutron star mergers, as we discuss in Section \ref{tde}. Finally, we summarize our findings in Section \ref{disc}, where we also comment on the detection prospects of late afterglows at $0.1-10$ GHz  with EVLA and LOFAR.

\section{The Non-Relativistic Afterglow Stages}\label{results}

\subsection{Analytical Estimates}\label{analytic}
In this section we use analytical arguments to calculate the GRB light curve in the non-relativistic stages, and its dependence on the various model parameters of the GRB blast wave. Our analytical estimates are verified and calibrated by the  
numerical results  of Section \ref{numerics}.

Let us consider a jet with (beaming-corrected) energy $E_{j}=10^{51.5}E_{51.5}\unit{erg}$ propagating in a  uniform medium\footnote{In Appendix \ref{app:A} we perform a similar analysis, for a wind profile.}  with number density $n=10\,n_{1}\unit{cm^{-3}}$. According to the analytical study by \citet{wygoda11} and the relativistic hydrodynamic simulations by \citet{ve12b}, the spherical ST solution can be used to estimate the observed flux for on-axis observers at a time
\be\label{eq:tst}
t\gtrsim t\ditto{ST}\simeq200\, (E_{51.5}/n_1)^{1/3} \unit{days}~,
\ee
which typically follows the deceleration of the blast wave down to non-relativistic speeds \citep{piran04,zhang09}.\footnote{In the following, we neglect the effect of the cosmological redshift, since the late afterglow emission is observable only in relatively nearby ($z\lesssim0.5$) bursts, where redshift corrections are small.} After the transition to the ST solution, the blast wave radius evolves in time as 
\be
R\simeq7.5\times10^{17} (E_{51.5}/n_1)^{1/5}t_{\rm yr}^{2/5} \unit{cm}~,
\ee
and its velocity in units of the speed of light is
\be
\beta\simeq0.3\, (E_{51.5}/n_1)^{1/5}t_{\rm yr}^{-3/5}~,
\ee
where $t_{\rm yr}$ is the observer time in years. If the shock-accelerated electrons populate a power law of index $p$ containing a fraction $\epsilon_e$ of the shock-dissipated energy, then the minimum electron Lorentz factor will be
\be\label{eq:gmin}
\gamma_m-1\simeq\frac{p-2}{p-1}\epsilon_e\frac{m_p\,\beta^2}{2\,m_e}\equiv\frac{1}{8}\bar{\epsilon}_e \frac{m_p\beta^2}{m_e} ~,
\ee
where for compactness we have set $\bar{\epsilon}_e\equiv\epsilon_e\, 4(p-2)/(p-1)$. The ``deep Newtonian phase'' will start at the time $t\ditto{DN}$ when $(\gamma_m-1) m_e c^2\sim m_e c^2$, i.e., the bulk of the shock-accelerated electrons turn non-relativistic. This gives
\be\label{eq:tdn}
t\ditto{DN}\simeq1.5 \,(E_{51.5}/n_1)^{1/3}\bar{\epsilon}_{e,-1}^{5/6} \unit{years}~,
\ee
where $\bar\epsilon_{e}=0.1\,\bar\epsilon_{e,-1}$. This estimate for $t\ditto{DN}$ is in good agreement with the numerical results of Section \ref{numerics}. 

We now estimate the synchrotron flux expected in the radio band.  Let us parameterize the post-shock  magnetic energy density as being a fraction $\epsilon_{\scriptsize{\rm B}}=0.01\,\epsilon_{\scriptsize{\rm B},-2}$ of the thermal energy density of the shocked fluid $e_{th}\sim 2\, n m_p \beta^2 c^2$, so that the  field strength is
\be
B\simeq0.04\, \epsilon_{\scriptsize{\rm B},-2}^{1/2}E_{51.5}^{1/5}n_1^{3/10}t_{\rm yr}^{-3/5} \unit{G}~.
\ee
When the low-energy end of the electron distribution is still ultra-relativistic (i.e., in the limit  by \citet{frail00}), the characteristic synchrotron frequency $\nu_{\rm pk}\propto \gamma_{\rm pk}^2B$ will be emitted by electrons with $\gamma_{\rm pk}\sim\gamma_m$, so
\be
\nu_{\rm pk}\sim\nu_m\simeq10^6\, \bar{\epsilon}_{e,-1}^2\epsb^{1/2}E_{51.5}\,n_1^{-1/2}t_{\rm yr}^{-3} \unit{Hz}~.
\ee
The overall luminosity $L_{\rm pk }\sim L_m$ at the peak frequency is the product of  the total number of radiating electrons ($\propto R^3 n$) and the peak power of a single electron ($\propto B$). Using the definition of $\gamma_m$ in \eq{gmin}, we can alternatively write that $L_{\rm pk}\propto B\,R^3\bar{\epsilon}_e e_{th}/(\gamma_m-1) $, an expression that will be useful below. We obtain
\be
\!\!\!\!\!L_{\rm pk}\!\sim\! L_m\simeq3\times10^{32} \, \epsb^{1/2}E_{51.5}^{4/5}n_1^{7/10} t_{\rm yr}^{3/5}\unit{erg\,s^{-1}\,Hz^{-1}}
\ee
and the flux observed at $1\,$GHz (for a burst with luminosity distance $d_L\simeq 10^{27.5}\,d_{27.5}\unit{cm}$, as appropriate for GRB 030329) is
\be\label{eq:fnu_frail}
\!\!\!\!\!\!F_\nu&=&L_m/(4 \pi d_L^2)\, (\nu_m/\nu)^{(p-1)/2}~\nonumber\\
\!\!\!\!\!\!&\simeq&0.3\,  \bar{\epsilon}_{e,-1}^{p-1}\epsb^{\frac{1+p}{4}}E_{51.5}^{\frac{3+5p}{10}} n_1^{\frac{19-5p}{20}} t_{\rm yr}^{\frac{3(7-5p)}{10}} \nu_{\rm GHz}^{\frac{1-p}{2}} \,d_{27.5}^{-2} \unit{mJy}~~
\ee
which has been rescaled down by a factor of $\sim4$ to match the numerical results obtained in Section \ref{numerics} at $t=1000$ days. The last expression assumes optically thin emission of slow-cooling electrons in the observed band. This is typically the case at $\gtrsim$GHz frequencies and years after the burst (we refer to Appendix \ref{app:B} for a discussion of the role of synchrotron self-absorption).   

\subsubsection {The Deep Newtonian Regime}
The  analysis above applies to the limit discussed by \citet{frail00}, i.e., the blast wave is non-relativistic, but the distribution of shock-heated electrons has $\gamma_{ m}\gg 1$. 
The applicability of this regime is rather narrow. Independently of  the choice of $\epsilon_e$, the electrons eventually turn non-relativistic, and the previous afterglow treatment becomes invalid. Moreover, as shown in Equations \eqn{tst} and \eqn{tdn}, in the case $\bar{\epsilon}_e\lesssim0.03$ the electrons become non-relativistic before the blast wave relaxes to the ST solution, so the non-relativistic stage discussed by \citet{frail00} does not occur.

 While in the  limit presented by \citet{frail00} most of the electron energy is contributed by particles with $\gamma_{\rm pk}\sim\gamma_{m}$, in the deep Newtonian phase the energy census is dominated by electrons with $\gamma_{\rm pk}\sim2$. This follows from  the theory of Fermi acceleration in shocks \citep[e.g.,][]{blandford_ostriker_78,bell_78,blandford_eichler_87}. The reason is that the spectrum of  accelerated particles follows a power-law distribution in {\it momentum} with slope $p$. In units of the dimensionless kinetic energy $x=\gamma-1$, we obtain $dN/dx\propto(x^2+2x)^{-(p+1)/2}(1+x)$, for $x\geq\gamma_m-1$. For $\gamma_m\gg1$, one recovers the familiar ultra-relativistic result, with the energy census  dominated by electrons having $\gamma_{\rm pk}\sim\gamma_m$. In contrast, for $\gamma_m-1\ll1$ and $2<p<3$, the energy of the electron distribution (i.e., $x^2 dN/dx$) peaks at $x_{\rm pk}\sim1$, or $\gamma_{\rm pk}\sim2$.\footnote{When $p\geq3$ our treatment is not applicable since $\gamma_m$ dominates both in energy and particle number (as in the ultra-relativistic case). Since theory and  observations suggest that $p<3$, in the main text we only explore the case $2<p<3$, and we defer to Appendix \ref{app:C} the treatment of steeper particle distributions.}
 
 
In the deep Newtonian phase, the flux declines as $F_\nu\propto t^{-3(1+p)/10}$, due to the following argument. The peak synchrotron frequency $\nu_{\rm pk}$ will be emitted by the electrons contributing most of the energy, i.e., with $\gamma_{\rm pk}\sim2$. So, the peak frequency in the deep Newtonian phase is
\be\label{eq:nupk_dn}
\nu_{\rm pk}\simeq4\times 10^5\,  \epsilon_{\scriptsize{\rm B},-2}^{1/2}E_{51.5}^{1/5}n_1^{3/10}t_{\rm yr}^{-3/5} \unit{Hz}~.
\ee
The luminosity emitted at the peak frequency scales as  $L_{\rm pk}\propto B R^3  \bar{\epsilon}_e e_{th}/\gamma_{\rm pk}$, which is the same expression as before, but with $\gamma_m$ replaced by $\gamma_{\rm pk}$. We obtain $L_{\rm pk}\propto t^{-3/5}$. So,
the flux at the frequency $\nu$ scales as
\be
F_\nu=L_{\rm pk}/(4 \pi d_L^2)\, (\nu_{\rm pk}/\nu)^{(p-1)/2}\propto t^{\frac{-3(1+p)}{10}} \nu^{\frac{1-p}{2}}
\ee
By matching this temporal decay with the flux in \eq{fnu_frail} computed at the transition time $t=t\ditto{DN}$, we find the flux in the deep Newtonian phase to be
\be\label{eq:fnu_dn}
\!\!\!\!\!\!F_\nu\!=\!0.2\,  \bar{\epsilon}_{e,-1}\epsb^{\frac{1+p}{4}}E_{51.5}^{\frac{11+p}{10}} n_1^{\frac{3+3p}{20}}t_{\rm yr}^{\frac{-3(1+p)}{10}} \nu_{\rm GHz}^{\frac{1-p}{2}}\, d_{27.5}^{-2} \!\unit{mJy}~~~
\ee
The numerical value is calibrated by the simulation results of Section \ref{numerics} at $t=1000$ days, for a burst with luminosity distance $d_L\simeq 10^{27.5}\unit{cm}$, as appropriate for GRB 030329. The simulations also verify the dependence of the flux on the various model parameters
as predicted by \eq{fnu_dn}.

Finally, we point out that the scalings derived above can be recovered in the formalism appropriate to the ultra-relativistic case $\gamma_m\gg1$, modulo a few changes. When $\gamma_m$ falls below unity, i.e., at the onset of the deep Newtonian phase, one should set $\gamma_m=1$, and account for the fact that only a small fraction of the electrons are now ultra-relativistic. If the fraction of accelerated electrons is constant and equal to $\zeta_{e,\rm UR}$ when $\gamma_m\gg1$ (in this work, we assume $\zeta_{e,\rm UR}=1$ for simplicity), during the deep Newtonian phase it is sufficient to replace $\zeta_{e,\rm UR}$ by $\zeta_{e,\rm DN}=\zeta_{e,\rm UR}\frac{p-2}{p-1}\,\epsilon_e\, (\Gamma-1) \,m_p/m_e\ll1$, which evolves as $\zeta_{e,\rm DN}\propto \beta^2\propto t^{-6/5}$.

\subsection{Synthetic Light Curves from Relativistic Hydrodynamical Simulations}\label{numerics}
Our analytical estimates are based on a simple model where the blast wave is assumed to relax into a spherical ST solution at its
non-relativistic stages. Here we demostrate that such a model gives a fairly good description of the observed properties of the 
blast wave, by comparing our analytical estimates to the results of relativistic hydrodynamical simulations.

\subsubsection{The New Prescription for Particle Acceleration}
We have computed synthetic light curves of GRB afterglows in the radio band by employing the \textsf{Afterglow Library} described in \citet{ve12} \cite[see also][]{zhang09,ve09,ve10,ve10b,ve11}.\footnote{The Afterglow Library is publicly available at {\sf http:/$\!$/cosmo.nyu.edu/afterglowlibrary}.} The library calculates the full light curves and spectra using linear radiative transfer (including synchrotron self-absorption) through snapshots of relativistic hydrodynamical simulations of GRB jets.

The hydrodynamical simulations employed by \citet{ve12} can follow the GRB jet from the ultra-relativistic phase down to non-relativistic velocities. When computing the radiative signature of GRB afterglows, \citet{ve12} implicitly assumed that the  distribution of  emitting electrons is a power law in energy of the form $dN/d\gamma\propto \gamma^{-p}$ for $\gamma\geq\gamma_m=\frac{p-2}{p-1}\epsilon_e e_{th,loc}/(n_{loc} m_e c^2)$, where $e_{th,loc}$ and $n_{loc}$ are the thermal energy density and the number density of the local fluid.\footnote{So far, we have used $e_{th}$ for the thermal energy density of the material right behind the shock, whereas $n$ was the number density of the circumburst medium.} As we have discussed in Section \ref{analytic}, such a parametrization is not appropriate for $t\gtrsim t\ditto{DN}$, when the bulk of the accelerated electrons become non-relativistic (and $\gamma_m$ as defined above falls below unity).


We have modified the radiation module of the \textsf{Afterglow Library} to account for the transition to the deep Newtonian regime. When $\gamma_m$ (as defined above) falls below unity, we set $\gamma_m=1$.\footnote{We have tested that a different choice for the threshold value (yet, still of order unity) does not appreciably change the light curves in the deep Newtonian phase.} Also, the emission and absorption coefficients are taken to be proportional to $(p-2)\epsilon_e e_{th,loc}/(\gamma_m m_e c^2)$, which is valid both for ultra-relativistic electrons [it reduces to $(p-1) n_{loc}$, when $\gamma_m=\frac{p-2}{p-1}\epsilon_e e_{th,loc}/(n_{loc} m_e c^2)\gg1$] and in the deep Newtonian regime (with $\gamma_m=1$). 

In principle, the transition to the Newtonian regime could be accompanied by several other changes in the distribution of accelerated electrons. For example, the power-law slope might decrease from the universal prediction $p\simeq2.23$ of Fermi acceleration in relativistic shocks \citep{kirk_00,keshet_waxman_05} to the standard value $p\simeq2$ expected in non-relativistic shocks \citep[e.g.,][]{blandford_eichler_87}. Or, the fraction $\epsilon_e$ of shock energy transferred to the accelerated electrons might drop by one or two orders of magnitude, as the blast wave becomes non-relativistic. However, the magnitude of such changes is still a subject of active research, and we do not implement them in our radiation module. In contrast, the evolution of the bulk electrons down to non-relativistic energies is an \textit{inevitable} outcome of the deceleration of the blast wave at $t\gtrsim t\ditto{DN}$. We now show how this affects the late-time light curves of GRB radio afterglows.

\begin{figure}[tbp]
\begin{center}
\includegraphics[width=0.5\textwidth]{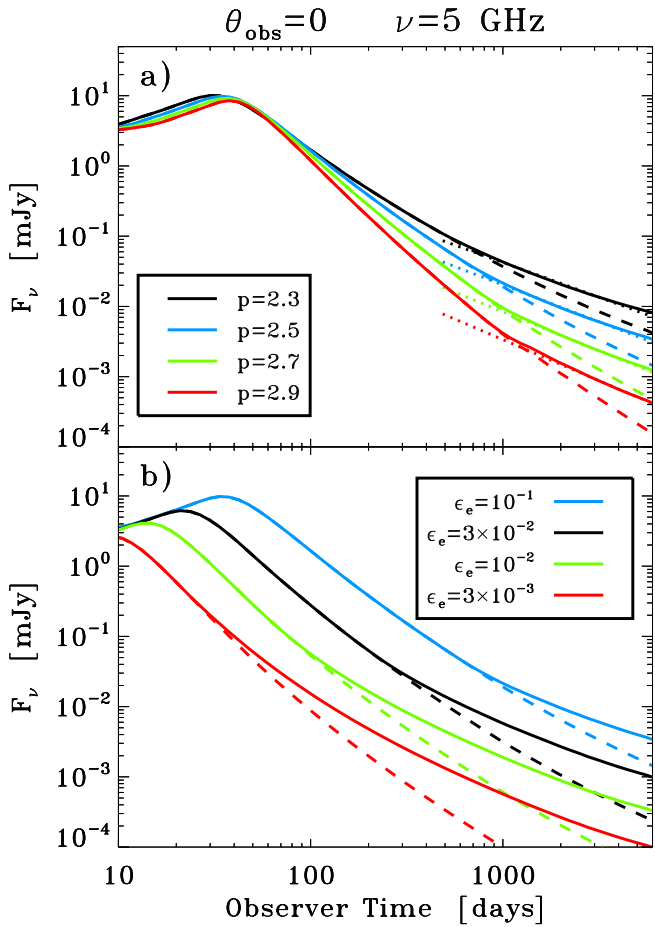}
\caption{Synthetic light curves extracted from relativistic hydrodynamical simulations, for on-axis observers (i.e., $\theta_{\rm obs}=0$) at frequency $\nu_{\rm obs}=5\unit{GHz}$. We employ $E_{51.5}=1$, $n_1=1$, $\epsb=1$ and the luminosity distance $d_L\simeq10^{27.5}\unit{cm}$ appropriate for GRB 030329. In both panels, we show both the limit discussed by \citet[][dashed lines]{frail00} and  the deep Newtonian stage presented in this work (solid lines). Top panel: for fixed $\epsilon_{e}=0.1$, we show the light curves for different electron power-law slopes $p$. The contribution of the counter-jet is artificially neglected, to show clearly that the late-time decay is in agreement with our analytical estimates (dotted lines). Bottom panel: for fixed $p=2.5$, we show the light curves for different $\epsilon_e$, still neglecting the contribution of the counter-jet.}
\label{fig:fig1}
\end{center}
\end{figure}

\begin{figure}[tbp]
\begin{center}
\includegraphics[width=0.5\textwidth]{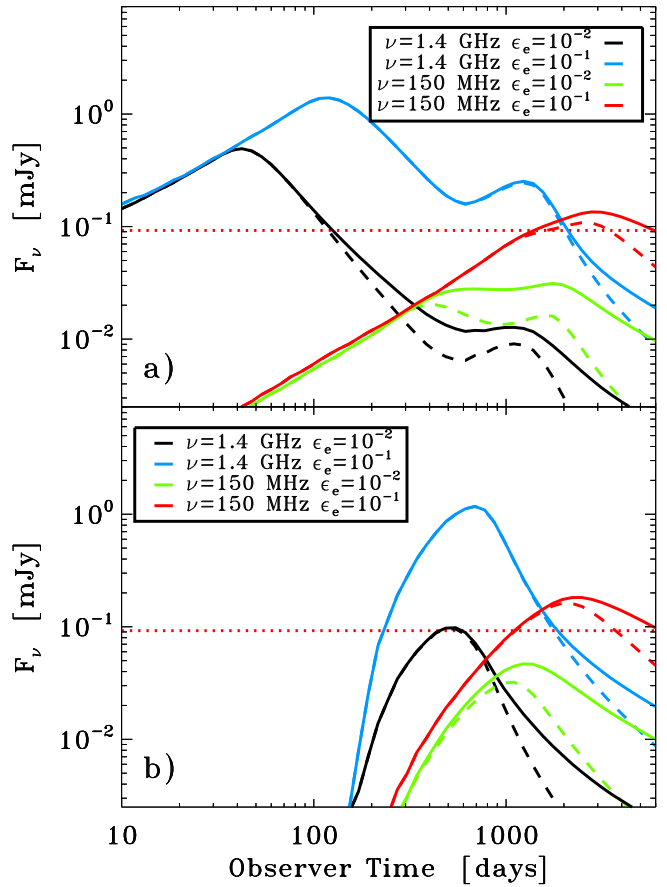}
\caption{Synthetic light curves extracted from relativistic hydrodynamical simulations, for frequencies $150\unit{MHz}$ and $1.4\unit{GHz}$ and electron acceleration efficiencies $\epsilon_e=10^{-2}$ and $\epsilon_e=10^{-1}$. We employ $E_{51.5}=1$, $n_1=1$, $\epsb=1$, $p=2.5$ and  the luminosity distance $d_L\simeq10^{27.5}\unit{cm}$ appropriate for GRB 030329. In both panels, we show both the limit discussed by \citet[][dashed lines]{frail00} and  the deep Newtonian stage presented in this work (solid lines). The rebrightening at late times is produced by the counter-jet. Top panel: on-axis observer ($\theta_{\rm obs}=0$). Bottom panel: off-axis observer ($\theta_{\rm obs}=\pi/2$). In both panels, the red dotted horizontal lines show the sensitivity limit of LOFAR at $150\unit{MHz}$ with 24 hours of integration time.}
\label{fig:fig2}
\end{center}
\end{figure}

\subsubsection{Results}\label{res}
In \fig{fig1}, we show GRB light curves at 5 GHz for on-axis observers, computed using the  \textsf{Afterglow Library} described in \citet{ve12}. We neglect the contribution of the counter-jet,  to emphasize how our results differ from the prescription of electron acceleration employed by earlier studies. 

In the top panel, we show that the late-time evolution (solid lines) is well described by a power law, with the slope predicted by \eq{fnu_dn} (dotted lines). The temporal decay in our model  (solid lines)
 is always shallower than in the regime discussed by \citet[dashed lines]{frail00}, in agreement with the analytical estimates in \eq{fnu_frail} and \eq{fnu_dn}. As the electron power-law slope $p$ varies from $2$ to $3$, the difference between the two temporal decay slopes increases. For $p=3$, at the upper extreme of the range $2<p<3$ where our analysis is applicable, $F_\nu\propto t^{-1.2}$ in our model, whereas $F_\nu\propto t^{-2.4}$ in \citet{frail00}.
 
 The light curves plotted in the top panel are obtained for a fixed fraction $\epsilon_e=0.1$ of shock energy transferred to the accelerated electrons. Yet, the power-law slope $p$ changes, as indicated in the legend, so $\bar{\epsilon}_e=\epsilon_e\, 4 (p-2)/(p-1)$ is larger for higher values of $p$. This causes the onset of the deep Newtonian phase (as marked by the divergence between the solid and dashed lines) to occur later for larger $p$, in agreement with \eq{tdn}. A more dramatic effect on the onset time $t_{\rm DN}$ of the deep Newtonian phase is produced by a change in the electron acceleration efficiency $\epsilon_e$, as shown in the bottom panel of \fig{fig1}. As stated by \eq{tdn}, the transition to the deep Newtonian regime happens earlier for smaller $\epsilon_e$. For $\epsilon_e\lesssim10^{-2}$, the deep Newtonian phase promptly follows the relativistic deceleration stage, and the non-relativistic phase of \citet{frail00} does not occur.
 
The effect of the electron acceleration efficiency $\epsilon_e$ on the afterglow flux is further investigated in \fig{fig2}, where we show the light curves obtained with our model (solid lines), and the results following \citet{frail00} (dashed curves). We include the contribution of the counter-jet, and 
we explore two values of  the electron acceleration efficiency ($\epsilon_e=10^{-2}$ and $\epsilon_e=10^{-1}$), two observing frequencies ($\nu=150$ MHz and $\nu=1.4$ GHz ), and two choices for the line of sight (on axis observer in the top panel and edge-on observer in the bottom panel). 

At $\nu=1.4$ GHz, if the electron acceleration efficiency is small (see the black line in the top panel), the on-axis flux from the forward shock at $t\sim1000$ days is so bright  (as compared to predictions based on \citealt{frail00}) that it might mask the contribution from the counter-jet. In fact, the counter-jet rebrightening appears much shallower in our model (solid black line in the top panel) than in the prescription by \citet[][dashed black curve]{frail00}. This might explain the absence of clear counter-jet signatures in the late-time radio afterglow of GRB 030329 \citep{mesler12b}. Yet, this would require a relatively small electron acceleration efficiency ($\epsilon_e\lesssim0.01$). In fact, for the fiducial value $\epsilon_e=0.1$ (blue lines in the top panel), the counter-jet peak flux at $t\sim1000$ days does not depend on the prescription for electron acceleration. It is only at later times ($t\gtrsim2000$ days) that the flux approaches the predictions for the deep Newtonian phase as discussed in \eq{fnu_dn}, modulo an increase by a factor of two for the counter-jet contribution.

At smaller frequencies ($\nu=150$ MHz, green lines for $\epsilon_e=10^{-2}$ and red for $\epsilon_e=10^{-1}$), \fig{fig2} shows that even the peak of the light curve is sensitive to the prescription of electron acceleration. Our formalism (solid lines) predicts a broader peak with higher flux, as compared to the model by \citet{frail00} (dashed lines).\footnote{For the green lines in the top panel, the peak at $t\sim500$ days is due to the self-absorption frequency passing through the observing band, whereas the bump at $t\sim2000$ days is produced by the contribution of the counter-jet.} This has two important consequences. First, it will improve the chances for LOFAR detections of GRB afterglows at late times \citep{horst08}, since the peak flux for on-axis observers (red line in the top panel) approaches the sensitivity limit of LOFAR at 150 MHz \citep{lofar13}, with 24 hours of integration time (red dotted line). Second, the estimates relative to the detection at $\sim100$ MHz frequencies of off-axis afterglows (orphan afterglows) should be revised upwards, if the blast wave is in the deep Newtonian phase. In this stage, the peak flux for edge-on observers  (red line in the bottom panel)  is within the sensitivity window of LOFAR at $150$ MHz.


\section{Other Applications: Afterglows from TDE Jets, Radio Supenovae, Slow Ejecta from
Neutron Star Mergers}\label{tde}
The theory developed here for the emission from the late stages of a GRB blast wave 
is applicable to a number of sources with trans-relativistic ($\beta\sim 0.1-0.5$) ejecta.
Our prediction that the observed flux density will be larger than argued by previous models \citep[e.g.,][]{frail00}, as a result of an unavoidable change in the properties of the shock-accelerated electrons, indicates that the detectability of such sources by current 
and future observing facilities is better than previously thought.

The recently discovered transient jet from a supermassive black hole \citep{levan11}
is believed to be produced by a stellar tidal disruption event \citep[TDE;][]{bloom11,burrows11,zauderer11}. The radio emission from this event is modeled as the result
of the interaction between the jet and the interstellar medium, similarly to GRB afterglows \citep[e.g.,][]{giannios11,metzger12,berger12,zauderer13}. The radio emission remains bright even two years after the TDE \citep{zauderer13} and should 
remain observable at $\sim$GHz frequencies for years to come. Given the estimates of 
the blast energy and surrounding density \citep[][]{zauderer13}, 
we expect that the blast may enter the deep Newtonian regime in the next few years, 
making a good target to test our model.   
  
Trans-relativistic explosions have been observationally inferred in a number of supenovae \citep[e.g.,][]{soderberg08}
and are theoretically predicted in double neutron star mergers \citep[kilonovae;][]{li98,metzger10,barnes13}. Electromagnetic signatures
of the latter is critical in maximizing the science outputs from future gravitational wave detections \citep[e.g.,][]{metzger12b}. Still, 
previous estimates for the radio emission from such transients were rather pessimistic
for the detection prospects, if the ejecta have velocity $\beta \simless 0.2$ \citep{nakar11}.
Following our prescription for the electron distribution, the peak flux $f_{\rm peak}$ for ejecta with $\beta \sim 0.1$
 is typically a factor of a few higher than that calculated using the older 
prescription for electron acceleration (for $\epsilon_e=0.1 $ and $p=2.5$).
As a result of the increased flux, the timescale $T$ over which the source is detectable is also prolonged.
Since the detectability of a source in a magnitude-limited survey scales as $\sim f_{\rm peak}^{3/2}T$, there will be a substantial increase 
in the detection rates.

\section{Summary and Discussion}\label{disc}
We have described a novel evolutionary stage of GRB afterglows, when the bulk of the shock-accelerated electrons are non-relativistic, and most of the electron energy is contributed by particles with $\gamma\sim2$. This phase, which we refer to as the ``deep Newtonian phase,'' necessarily regulates the late-time evolution of GRB afterglows, at  $t\gtrsim t_{\rm DN}\simeq 3\, \epsilon_{e,-1}^{5/6} t_{\rm ST}$, where $t_{\rm ST}$ marks the transition to the non-relativistic spherically-symmetric Sedov--Taylor solution. For typical parameters, the onset of the deep Newtonian stage occurs $\sim0.5-$several years after the GRB.  The deep Newtonian phase usually follows the non-relativistic regime discussed by \citet{frail00}, which applies if the blast wave is non-relativistic, but the accelerated electrons are still ultra-relativistic. However, if the electron acceleration efficiency is $\epsilon_{e}\lesssim0.03$, then $t_{\rm DN}\lesssim t_{\rm ST}$, i.e., the blast wave transitions from the relativistic deceleration stage directly to the deep Newtonian phase, and the non-relativistic stage presented by \citet{frail00} does not occur.

We now describe the observational implications of our findings. Our main focus is on GRB afterglows, but our conclusions will be relevant for other systems with trans-relativistic outflows, i.e., afterglows from tidal disruption jets, radio supernovae, and trans-relativistic ejecta from double neutron star mergers.

Radio observations at $1-10$ GHz for $t\gtrsim 500-1000$ days are available for a number of GRBs \citep[e.g.,][]{frail04}, most notably GRB 970508 \citep{waxman98}, GRB 980703 \citep{berger04}, and GRB 030329 \citep{frail05,mesler12b}. The shallow $\sim t^{-1}$ decline of a number of late radio afterglows \citep[e.g.,][]{frail04} may indicate that the deep Newtonian phase has been already observed in several bursts (we predict that the flux  in the deep Newtonian phase should indeed decay as $ F_\nu\propto t^{-(0.9\div1.2)}\sim t^{-1}$). Even more importantly, the current brightness and the fact that the flux drops as $\sim t^{-1}$ make it possible to continue studying these afterglows for several years to come. With the full frequency coverage of the Expanded VLA (EVLA), it will  be possible to cover the entire $1-10$ GHz frequency range in a few hours of observations, to a sensitivity that is about an order of magnitude better than that of the VLA. 

In summary, current and future observational capabilities will open a new window on the late-time stages of GRB afterglows. A single radio measurement (single epoch and frequency) will probably be insufficient to determine whether the blast wave is in the deep Newtonian phase, or in the shallow Newtonian regime discussed by \citet{frail00}. Our model predicts brighter fluxes, since $F_\nu \propto t^{-3(p+1)/10}\propto t^{-(0.9\div1.2)}$ in the deep Newtonian phase (see \eq{fnu_dn}), whereas $F_\nu  \propto  t^{-3(5p-7)/10}\propto t^{-(0.9\div2.4)}$ in the regime of \citet{frail00} (see \eq{fnu_frail}). Yet, the flux difference might be compensated by a larger value for the fitting parameter $\epsilon_e$. If observations at two radio frequencies are available, the radio spectral slope $-(p-1)/2$ will constrain the electron power-law slope $p$, and the temporal decay of the late-time light curve will promptly distinguish between shallow and deep Newtonian regime.

Radio observations of late-time afterglows will provide an energy estimate of the explosion \citep[e.g.,][]{berger04,shivvers11}. The calorimetric estimates in the Newtonian phase would be independent of jet collimation, since the blast wave approaches spherical symmetry. Second, this regime relies on the simple and well-understood ST dynamics of spherical blast waves. Finally, the afterglow is observable for several hundred days in the radio band, which allows better constraints on the calorimetry. Even with a single radio measurement, one can put some constraints on the burst energy: as described by \eq{fnu_dn},
one  fixes $\epsilon_{\rm B}, \epsilon_e$ and the only free parameters are $n, E_j$. In the deep Newtonian regime discussed in this work, we have the
advantage that $ F_\nu\propto t^{-(0.9\div1.2)}\sim t^{-1}$ regardless of the uncertain slope $p$, whereas in the  limit  of \citet{frail00} the temporal decay depends more sensitively on $p$.

Finally, the deep Newtonian phase will directly impact the detection prospects at the frequencies probed by LOFAR \citep{horst08,lofar13}. For $\sim $GHz frequencies, the peak of the light curve happens before the deep Newtonian regime. However, as shown in \fig{fig2}, for LOFAR frequencies ($\sim100-200$ MHz), our model for the deep Newtonian phase gives a brighter and broader peak, with respect to the prescription discussed by \citet{frail00}. This will significantly improve the chances for LOFAR detections of GRB afterglows at late times, for both on-axis observers and edge-on observers (i.e., orphan afterglows).

\acknowledgements
L.S. is supported by NASA through Einstein
Postdoctoral Fellowship grant number PF1-120090 awarded by the Chandra
X-ray Center, which is operated by the Smithsonian Astrophysical
Observatory for NASA under contract NAS8-03060. L.S. gratefully thanks H.~J. van Eerten for help in using the Afterglow Library.

\appendix
\section{The Case of Steep ($\lowercase{p}\geq3$) Electron Distributions}\label{app:C}
The results presented in the main text assume that the  distribution of accelerated electrons is in the form of 
a power law with slope $2<p<3$, for $\gamma\geq\gamma_m$. In this case, which is supported by the observations, most of the electrons (by number) are close to the minimum Lorentz factor $\gamma_m$, whereas the energy census is dominated by electrons with $\gamma\sim2$. Here, we derive the temporal scalings expected in the deep Newtonian phase for steep electron spectra with $p\geq 3$, when $\gamma_m$ dominates both in energy and particle number.
 
If we define the characteristic frequency $\nu_{\rm pk}$ as being emitted by mildly relativistic particles with $\gamma_{\rm pk}\sim2$, we have $\nu_{\rm pk}\propto t^{-3/5}$ as in \eq{nupk_dn}. Yet, we remind that for $p\ge3$, mildly relativistic electrons with $\gamma\sim2$ do not dominate the energy balance, as opposed to the case $2<p<3$ presented in the main text. 

The flux emitted at the characteristic frequency $\nu_{\rm pk}$ needs to take into account that only a small fraction of electrons are accelerated to ultra-relativistic energies,  thus contributing to the emission. If $x_m=\gamma_m-1\propto \beta^2$ is the minimum of the electron distribution, then the fraction of electrons that are ultra-relativistic (i.e., with $\gamma\gtrsim2$) is $\sim x_m^{(p-1)/2}$. The flux at the characteristic frequency $\nu_{\rm pk}$ then scales as $L_{\rm pk}\propto B R^3 n \,x_m^{(p-1)/2}\propto t^{3(2-p)/5}$. Finally, the flux at the observing frequency $\nu$ decays in time as 
\be
F_\nu\propto L_{\rm pk}\, (\nu_{\rm pk}/\nu)^{(p-1)/2}\propto t^{\frac{3(5-3p)}{10}}\nu^{\frac{1-p}{2}},
\ee
which extends the result in \eq{fnu_dn} to the case $p\geq3$.

\section{Scalings in a Wind-like Density Profile}\label{app:A}
In the main body of the paper, we have assumed that the circumburst density is uniform, which is supported by recent radio observations of late-time afterglows \citep[e.g.,][]{mesler12b}. For the sake of completeness, here we provide the expected scalings in the case that the circumburst density follows a wind-like profile, $n=A/(m_p R^2)$. If we define $A=10^{12} A_{12} \unit{g \, cm^{-1}}$, the radius of the Sedov-Taylor solution evolves as 
\be
R\simeq4.4\times10^{18} (E_{51.5}/A_{12})^{1/3}t_{\rm 10\,yr}^{2/3} \unit{cm}~,
\ee
where $t=10 \,t_{10\,\rm{yr}} \unit{years}$. The blast velocity in units of the speed of light is
\be
\beta\simeq0.3\, (E_{51.5}/A_{12})^{1/3}t_{10\,\rm yr}^{-1/3}~.
\ee
The deep Newtonian phase  begins when $(\gamma_m-1)m_e c^2\sim m_e c^2$, where $\gamma_m$ is  calculated from \eq{gmin}. We find
\be
t\ditto{DN}\simeq 20\,(E_{51.5}/A_{12})\,\bar{\epsilon}_{e,-1}^{3/2} \unit{years}~,
\ee
which is significantly longer, for standard parameters, than in the case of a uniform medium.

By assuming that a fraction $\epsilon_{\rm B}$ of the shock energy is converted into magnetic fields, the field strength will be 
\be
B\simeq0.002\, \epsilon_{\scriptsize{\rm B},-2}^{1/2}A_{12}^{1/2}t_{10\,\rm yr}^{-1} \unit{G}~.
\ee
When the low-energy end of the electron distribution is still ultra-relativistic (i.e., in the limit by \citealt{frail00}), the characteristic synchrotron frequency emitted by electrons with $\gamma_{\rm pk}\sim \gamma_m$ is
\be
\nu_{\rm pk}\sim\nu_m\simeq5.2\times10^4\, \bar{\epsilon}_{e,-1}^2\epsb^{1/2}E_{51.5}^{4/3}\,A_{12}^{-5/6}t_{\rm 10\,yr}^{-7/3} \unit{Hz}~,
\ee
the luminosity at the peak frequency is
\be
\!\!\!\!\!L_{\rm pk}\!\sim\! L_m\simeq2.7\times10^{31} \, \epsb^{1/2}E_{51.5}^{1/3}A_{12}^{7/6} t_{\rm 10\,yr}^{-1/3}\unit{erg\,s^{-1}\,Hz^{-1}}~,
\ee
and the flux observed from a burst at a distance of $d_L\simeq 10^{27.5}\unit{cm}$ will be
\be\label{eq:fnu2_frail}
\!\!\!\!\!\!F_\nu&=&L_m/(4 \pi d_L^2)\, (\nu_m/\nu)^{(p-1)/2}~\nonumber\\
\!\!\!\!\!\!&\simeq&0.02\,  \bar{\epsilon}_{e,-1}^{p-1}\epsb^{\frac{1+p}{4}}E_{51.5}^{\frac{4p-3}{3}} A_{12}^{\frac{19-5p}{12}} t_{\rm 10\,yr}^{\frac{5-7p}{6}} \nu_{\rm GHz}^{\frac{1-p}{2}} \,d_{27.5}^{-2} \unit{mJy}~,
\ee
which is in agreement with the results by \citet{livio_waxman00}.

Different scalings are expected in the deep Newtonian phase. Under the assumption that the electron power-law slope is $2<p<3$, the peak frequency is emitted by mildly relativistic electrons with $\gamma_{\rm pk}\sim2$, so
\be
\nu_{\rm pk}\simeq2.4\times 10^4\,  \epsilon_{\scriptsize{\rm B},-2}^{1/2}A_{12}^{1/2}t_{\rm 10\,yr}^{-1} \unit{Hz}~,
\ee
the peak luminosity scales in time as $L_{\rm pk}\propto t^{-1}$, and the flux at frequency $\nu$ decays as
\be\label{eq:fnu2_dn}
\!\!\!\!\!\!F_\nu&=&L_{\rm pk}/(4 \pi d_L^2)\, (\nu_{\rm pk}/\nu)^{(p-1)/2}~\nonumber\\
\!\!\!\!\!\!&\simeq&
0.015\,  \bar{\epsilon}_{e,-1}\epsb^{\frac{1+p}{4}}E_{51.5}A_{12}^{\frac{p+1}{4}}t_{\rm 10\,yr}^{\frac{-(1+p)}{2}} \nu_{\rm GHz}^{\frac{1-p}{2}}\, d_{27.5}^{-2} \!\unit{mJy}~,
\ee
where the numerical factor has been calibrated such that  it matches the flux in \eq{fnu2_frail} at the onset of the deep Newtonian phase.

\section{Synchrotron Self-Absorption}\label{app:B}
The results presented so far implicitly assume that the self-absorption frequency lies below the observing frequency during the deep Newtonian phase. However, for observations at $\sim100$ MHz, the self-absorption frequency will sweep across the observing band at late times, during the deep Newtonian regime.  We now derive the temporal evolution of the self-absorption frequency, for a constant density medium and a wind-like profile. As in the main body of the paper, we assume that the electron power-law slope is in the range $2<p<3$.

The self-absorption frequency $\nu_{a}$ is such that the optical depth $\tau_\nu(\nu_{a})\!\sim\! \alpha_\nu R\!\sim\!1$, where the absorption coefficient is 
\be
\alpha_\nu (\nu_a) \propto \frac{\epsilon_e e_{th} B}{\gamma_{\rm pk} \nu_a^2} \left(\frac{\nu_a}{\nu_{\rm pk}}\right)^{-p/2}~,
\ee
which is appropriate if $\nu_a\gg \nu_{\rm pk}$. In the expression above, $\gamma_{\rm pk}\sim2$, $e_{th}\propto n\beta^2$ and $B\propto \sqrt{\epsilon_{\rm B}n}\,\beta$, where $\beta$ is the flow velocity. We now differentiate between a constant density medium and a wind-like profile.

If the circumburst density is uniform, then $R\propto t^{2/5}$, $\beta\propto t^{-3/5}$, $n\propto t^0$ and the peak frequency $\nu_{\rm pk}\propto t^{-3/5}$ (see \eq{nupk_dn}), so we obtain that the self-absorption frequency scales as
\be
\nu_a\propto t^{-\frac{3p+14}{5(p+4)}}\;\;\;\rm{[uniform~medium]}
\ee
For a wind-like profile, we have that $R\propto t^{2/3}$, $\beta\propto t^{-1/3}$, $n\propto t^{-4/3}$ and the peak frequency $\nu_{\rm pk}\propto t^{-1}$ (see Appendix \ref{app:A}), so that the self-absorption frequency scales as 
\be
\nu_a\propto t^{-\frac{3p+14}{3(p+4)}}\;\;\;\rm{[wind]}
\ee
In both cases, we point out that the self-absorption frequency drops faster than $\nu_{\rm pk}$. 

\bibliography{radio}

\end{document}